\documentclass[11pt]{article}

\pdfoutput=1
\usepackage[numbers]{natbib}
\usepackage{graphicx}
\usepackage{siunitx}
\usepackage[version=3]{mhchem}
\usepackage{amsmath}

\usepackage[colorlinks, allcolors=blue, hyperfootnotes]{hyperref}

\title{Inverse sensitivity of plasmonic nanosensors at the single-molecule limit}
\author{Boris Barbour\\~\\Institut de Biologie de l'Ecole Normale Sup\'erieure\\CNRS UMR 8197\\INSERM U1024\\PSL Research University\\75005 Paris\\France\\~\\Email: \tt boris.barbour@ens.fr}

\begin{document}

\maketitle
\newpage
\begin{abstract}
\bf

Recent work using plasmonic nanosensors in a clinically relevant
detection assay reports extreme sensitivity based upon a mechanism
termed \emph{inverse sensitivity}, whereby reduction of substrate
concentration increases reaction rate, even at the single-molecule
limit. This near-hom\-\oe o\-pathic mechanism contradicts the law of mass
action.  The assay involves deposition of silver atoms upon gold
nanostars, changing their absorption spectrum. Multiple additional
aspects of the assay appear to be incompatible with settled chemical
knowledge, in particular the detection of tiny numbers of silver atoms
on a background of the classic `silver mirror reaction'. Finally, it
is estimated here that the reported spectral changes require some
\SI{2.5E11} times more silver atoms than are likely to be produced. It
is suggested that alternative explanations must be sought for the
original observations.

\end{abstract}
\newpage

\section{Introduction}

\begin{figure}
\begin{center}
\includegraphics{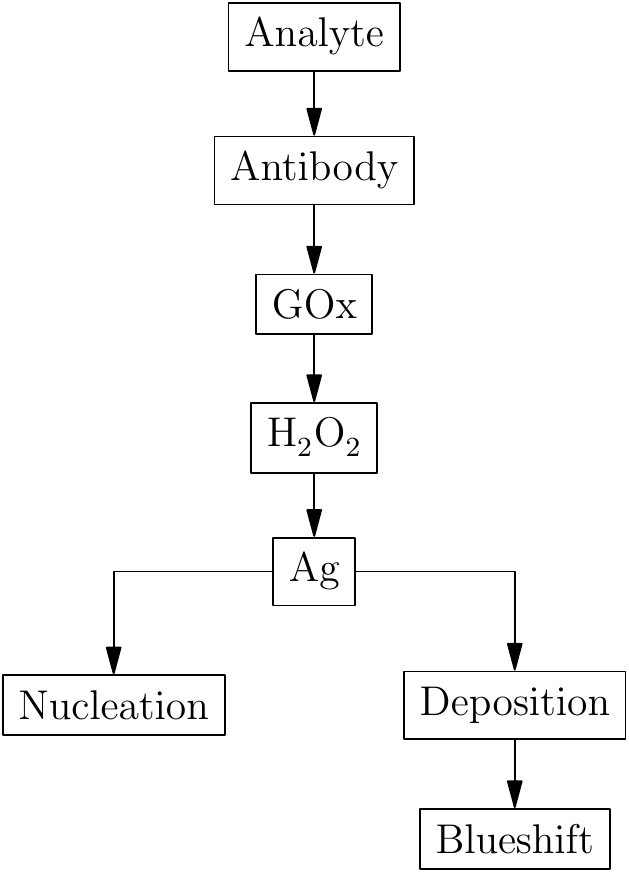}
\caption{{\bf Summary of assay reactions}. The silver atoms generated
  can either be deposited on gold nanostars (right branch), which
  leads to a blueshift of the absorbance spectrum, or can aggregate in
  free solution (`nucleation', left branch), in which case
  they do not affect the solution absorbance.}
\label{reaction}
\end{center}
\end{figure}

Rodriguez-Lorenzo et al.~\cite{Rodriguez-Lorenzo2012} report an ultra-sensitive method for
detecting analytes that can be recognised by an antibody. The PSA
protein is used to demonstrate the technique. The basis of the assay
is for the antigen to be recognised by antibodies conjugated with the
glucose oxidase enzyme (GOx), which then produces hydrogen
peroxide. The \ce{H_2O_2} in turn reduces silver ions, the resulting
silver atoms being deposited on gold nanoparticles (`nanostars'). The
deposition is detected by a blueshift of the absorption spectrum of
the solution of gold nanoparticles. The reactions are summarised in
Fig.~\ref{reaction} of this analysis.

\section{Inverse sensitivity}

Rodriguez-Lorenzo et al.~\cite{Rodriguez-Lorenzo2012} report bizarre, less-is-more reaction kinetics, according to which the reaction proceeds more quickly as the substrate concentration is reduced close to zero. In their own words (from the abstract of their paper):

\begin{quotation} \noindent \emph{However, because conventional
    transducers generate a signal that is directly proportional to the
    concentration of the target molecule, ultralow concentrations of
    the molecule result in variations in the physical properties of
    the sensor that are tiny, and therefore difficult to detect with
    confidence. Here we present a signal-generation mechanism that
    redefines the limit of detection of nanoparticle sensors by
    inducing a signal that is larger when the target molecule is less
    concentrated.}
\end{quotation}

\begin{figure}
\begin{center}
\includegraphics[width=\textwidth]{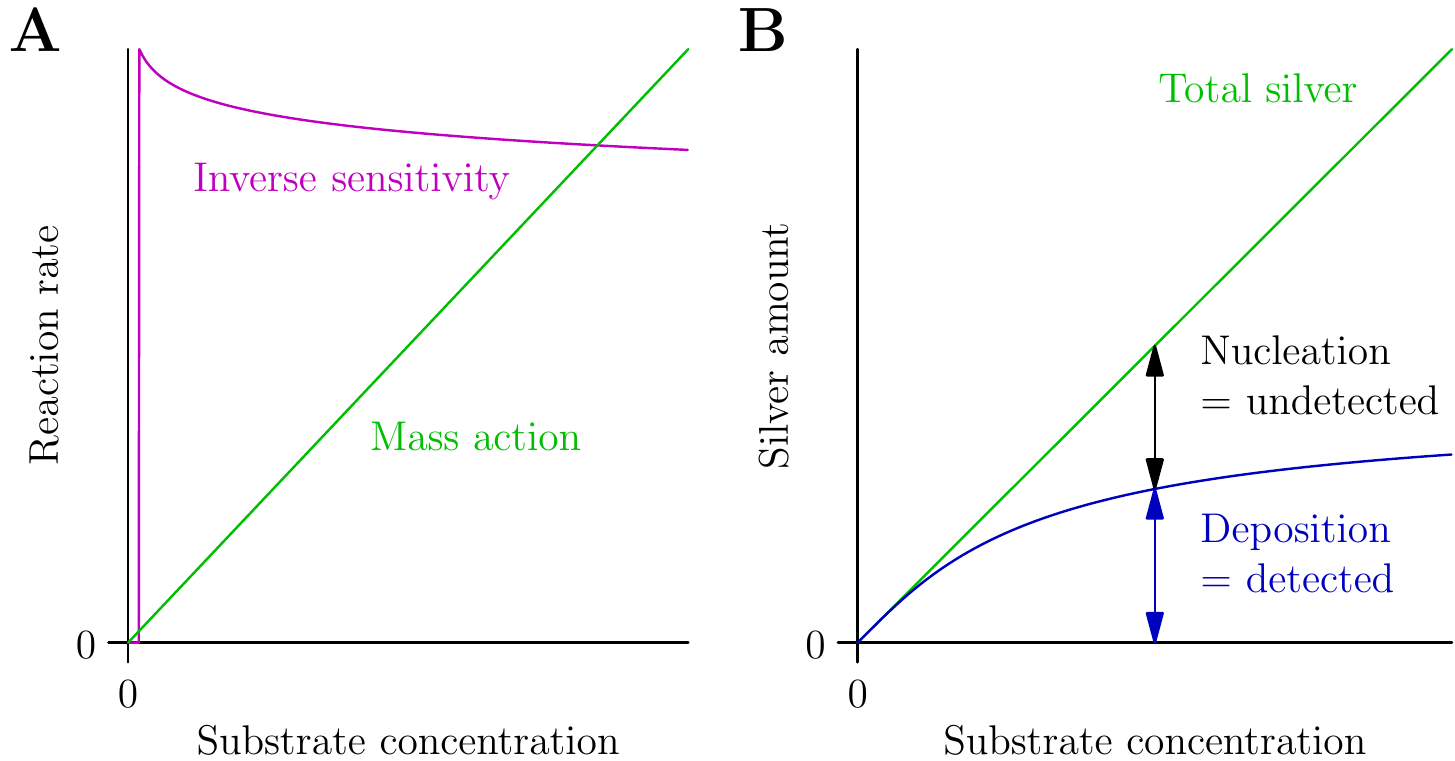}
\caption{{\bf Inverse sensitivity}. {\bf A}. The approximate form of reaction kinetics reported by Rodriguez-Lorenzo et al.~\cite{Rodriguez-Lorenzo2012} is depicted by the magenta curve (\emph{Inverse sensitivity}); the plot axes are linear. The reaction rate increases sharply at a very low threshold and then decays logarithmically with substrate concentration. In some experiments in the paper, the peak rate occurs at dilutions where \emph{less than one analyte molecule} is expected to be present. In contrast, according to the law of mass action, the rate should be proportional to the product of the substrate concentrations (or activities). {\bf B}. Whatever the substrate-dependence of silver production, the presence of a \emph{nucleation} reaction that consumes silver atoms in competition with the detected \emph{deposition} reaction can only \emph{reduce} the overall sensitivity of the reaction.}
\label{mass}
\end{center}
\end{figure}

The approximate form of the kinetics is sketched in Fig.~\ref{mass}A. As the substrate concentration is increased, the reaction rate rises abruptly from zero and then declines logarithmically (the authors' plots are semi-logarithmic) from a peak at extraordinarily low concentrations. In the GOx-detection experiment of their Fig.~1c, that peak occurs at a concentration where less than one molecule of GOx is expected to be present in the reaction volume (this is calculated in the next section). In contrast, the law of mass action states that the reaction rate is proportional to the product of the substrate concentrations (more accurately activities). Since only the analyte concentration is varied in the present experiments and it only appears with first-order kinetics, the reaction rate should simply be proportional to the analyte concentration at low concentrations\footnote{There is maybe some uncertainty regarding the dependence of the blueshift of the absorbance peak on the amount of silver deposited, but the former can be assumed to be an increasing function of the latter, so the conclusions reached here would be unaffected by the details of the relation.}. `Inverse sensitivity' appears to be spectacularly incompatible with the law of mass action.

The `explanation' offered by Rodriguez-Lorenzo et al.~\cite{Rodriguez-Lorenzo2012} for this
discrepancy is that spontaneous nucleation of pure silver
nanoparticles at high concentrations bypasses the deposition of silver
on the gold nanostars. Inspection of the assay reactions
(Fig.~\ref{reaction} of this analysis) shows that only the right-hand
branch, in which silver ions are deposited on the gold nanostars,
leads to the blueshift used to detect the analyte. Thus, the presence
of a competing nucleation reaction can only \emph{reduce} reaction
sensitivity (Fig.~\ref{mass}B of this analysis), irrespective of the
dependence of silver production, nucleation and deposition on analyte
concentration. Nucleation cannot increase assay
sensitivity\footnote{It is possible that in their conceptual argument
  the authors have confused the \emph{slope} of the analyte-blueshift
  curve, which could conceivably become negative at high silver
  concentrations, with the \emph{absolute} blueshift. In any case,
  what matters is the sensitivity at \emph{low} analyte
  concentrations, where the nucleation reaction is unlikely to
  proceed.}.

\section{Sensitivity and noise}

The assay is reported to have extraordinary sensitivity and exceptionally
low noise levels.

Fig.~2c of the paper (note that this and all figure references below
are to those in ref \cite{Rodriguez-Lorenzo2012}) reports the
detection of the difference between zero glucose oxidase and
\SI{1E-20}{g/ml} glucose oxidase, which represents an average of 0.04
molecules of GOx (MW = \SI{160}{kDa}) per ml. The precise reaction
volume is not reported in the paper, but would need to have been about
\SI{10}{ml} to have had a \SI{50}{\percent} chance of containing a
single molecule of GOx. A reaction volume of \SI{1}{ml} was used
elsewhere in the paper. As no statistics are given for this figure,
this observation may have been a lucky event whose replication was
never attempted.

Fig.~4 shows a quantification of the variability of the assay. In both
panels a and b, we see that \SI{1E-18}{g} PSA in the reaction volume
of \SI{1}{ml} is detected with fantastic precision compared to the
amount of \SI{1E-19}{g}. At both concentrations, the standard
deviation of the assay signal is in most cases smaller than the symbol
and in all cases smaller than a few percent of the maximum signal. But
\SI{1E-18}{g/ml} of PSA represents an average of just 23 molecules in
the reaction volume of 1\,ml. Such small quantities would necessarily
exhibit stochastic variation in the number of molecules present. By
Poisson statistics, 23 molecules should be associated with a standard
deviation of $\sqrt{23}$, equivalent to \SI{21}{\percent}. This
moreover represents a minimum. The signal amplification required to
detect such small quantities would certainly contribute additional
(high levels of) noise. Yet the authors consistently report improbably
low standard deviations.

This amazing sensitivity is at odds with a publication that predated
Rodriguez-Lorenzo et al. Li et al.~\cite{Li2011}, who used a variation
of the present assay to detect glucose (of which more below) with an
excess of GOx (as opposed to detecting GOx with excess glucose). Li et
al.~report a detection threshold of 10\,nM (although their Fig.~1
suggests that values in the micromolar range might be more
realistic). Even if a GOx molecule will obviously produce more silver
than a glucose molecule (estimated below), the difference between the
claimed detection thresholds for Li et al.~and Rodriguez-Lorenzo et
al.~is extreme: \SI{1E-8}{M} vs.~\SI{6E-23}{M}, a factor of
\SI{1.6E14}{}.

That single-molecule sensitivity is rendered even more unexpected by
another result in Li et al.~\cite{Li2011}. In their Fig.~S7, they
compare the abilities of glucose and \ce{H_2O_2} to reduce silver
ions. They report that \ce{H_2O_2} is much less effective. From this
we deduce that each molecule of \ce{H_2O_2} is by no means guaranteed
to reduce a silver ion. A poor yield at this stage of the assay would
reduce its sensitivity even further, making single-molecule detection
even more implausible.



\section{Silver mirror reaction}

Another problem is that the deposition of silver is triggered using a
mixture of \ce{AgNO_3} and \ce{NH_3}. The authors describe silver
being deposited on the gold nanoparticles (or aggregating via
nucleation and growth) as a result of reduction by the \ce{H_2O_2}
produced by glucose oxidase. In order for this to allow detection of
single molecules, a strict requirement is that absolutely no silver at
all be deposited in the absence of GOx and the \ce{H_2O_2} it
produces. However, it turns out that the assay reaction probably
contained two sources of reductants that were neither acknowledged nor
apparent in the results. Either of these sources would generate
background reductant concentration in excess of that arising during
the claimed detection of single analyte molecules.

The authors seem to have been unaware that they were using a classic
classroom reaction called the `silver mirror reaction'. The mixture of
\ce{AgNO_3} and \ce{NH_3} is called Tollen's reagent and is used to
detect aldehydes, whose presence triggers the deposition of a visually
impressive silver layer on any available surface. A nice description
of the reaction for motivating secondary school chemistry classes can
be found on the Royal Chemistry Society web site \cite{rsctollens}.
As demonstrated in that example, the reaction will produce a positive
in the presence of glucose, which has an aldehyde form in
solution. The problem is that in the assay of Rodriguez-Lorenzo et al.,
\SI{100}{mM} glucose is present as the substrate for glucose
oxidase. It seems inconceivable that it would not produce much more
silver deposition than the tiny amounts of \ce{H_2O_2} produced by a
few glucose oxidase molecules.

The paper by Li et al.~\cite{Li2011} provides direct support for our
assertion that glucose would reduce silver and generate a signal,
because they apply this assay precisely for the detection of glucose!
The \SI{100}{mM} glucose present in all experiments of
Rodriguez-Lorenzo et al.~would therefore generate a saturating
reduction of silver, against which background it would presumably be
impossible to detect single-molecule signals. In any case, these
expected and demonstrated background signals are simply absent from
the results reported by Rodriguez-Lorenzo et al.

The assay potentially contains a second source of reductant able to swamp
single-molecule signals. Luo et al.~\cite{Luo2010} report that gold
nanoparticles can catalyse the oxidation of glucose, producing
\ce{H_2O_2}. This catalysis is quite efficient for bare
nanoparticles. Some coatings of the gold can prevent the catalysis and
this may pertain in the experiments of Rodriguez-Lorenzo et
al. However, the covering would have to be perfect to allow
single-molecule detection.

\section{Nanoparticle numbers}

There are two further issues with quantitative aspects of the assay as
reported by the authors. I give a brief overview before
expounding the detailed arguments. 

The first problem is that the
quantities of enzyme involved will produce absolutely tiny amounts of
\ce{H_2O_2} and correspondingly tiny amounts of silver---enough to
deposit only a single atom on each of a very small fraction of the
gold nanoparticles present. It is extremely unlikely that addition of
a single atom will detectably change the absorbance spectrum of the
nanoparticle. 

The second and related problem is that the
expected large fraction of unmodified nanoparticles appears not to
contribute to the reported spectrum. Because the assay signal is the
absorbance of a dilute solution of nanoparticles, each nanoparticle
will contribute approximately independently to that absorbance. In the
absence of silver deposition, a control spectrum is obtained. Modified
nanoparticles would have a different spectrum depending on the degree
of modification. If a solution contains modified and unmodified
nanoparticles, a simple mixture of the two spectra should be
obtained. However, even under conditions where a very large fraction
of nanoparticles must have been unmodified, their dominant contribution to
the mixture spectrum was apparently absent.

The more detailed
explanations follow below and in the next section.

The assay is in two stages.  \ce{H_2O_2} is produced by the action of
GOx attached to the nanostars for 1 hour, then the silver ions are
added to trigger the silver deposition and/or nucleation, which are allowed
to proceed for another 2 hours. The precise reaction mixture for the
second stage is \SI{0.1}{mM} \ce{AgNO_3} + \SI{40}{mM} \ce{NH_3} added
to the \SI{10}{mM} MES buffer (pH 5.9) already present.

A first remark is that GOx is presumably totally inactivated by the
basic pH $\ge$ 10 of the second stage after addition of \ce{NH_3} (see Fig.~5 of
ref \cite{Wilson1992}). It also seems that GOx is strongly inhibited by
silver ions \cite{Nakamura1968}. So it is unnecessary to consider
\ce{H_2O_2} and silver produced next to the nanostars, just the
\ce{H_2O_2} concentration existing in the bulk solution at the end of
the first stage and the silver it produces during the second
stage. There is therefore no kinetic advantage in attaching the GOx to
the nanostars.

What is the concentration of \ce{H_2O_2}? The authors have omitted
details about the GOx used, so we'll assume it is the most active one
available from Sigma: G7141, with an activity of 100000--250000
units/g \cite{sigmagox}. The unit definition is: 

\begin{quotation}
\noindent 
One unit will
oxidize 1.0\,$\mu$mole of $\beta$-D-glucose to D-gluconolactone
and \ce{H_2O_2} per min at pH 5.1 at \SI{35}{\degreeCelsius}
equivalent to an \ce{O_2} uptake of 22.4\,$\mu$l/min. If the
reaction mixture is saturated with oxygen, the activity may increase
by up to \SI{100}{\percent}.
\end{quotation}

Another Sigma page
\cite{sigmaunitconditions} indicates that the final glucose
concentration under the conditions for the unit definition is
\SI{1.61}{\% w/v} or \SI{90}{mM}---similar to the \SI{100}{mM}
used by the authors.

Consider Fig.~2 and in particular the spectra in panel b for zero
glucose oxidase (black, blue) and \SI{1E-20}{g/ml} GOx (red). Using
the enzyme activity values just given, it can be calculated that this
low concentration of GOx would produce an \ce{H_2O_2} concentration of
\SI{1.5E-16}{M} after 1 hour. Generously assuming the production of
one silver atom per \ce{H_2O_2} molecule, \SI{9E4}{/ml} silver atoms
would be produced. (Above, we mentioned results that suggest that this
conversion is far from complete, which would result in many fewer
silver atoms.)

We now calculate the number of nanostars. The concentration of
nanostars is presumably the same as in the assays: [Au] =
\SI{0.25}{mM} (Methods). We'll also need the following values:
nanostar diameter \SI{60}{nm} (Fig.~2a; Methods), so radius
\SI{30}{nm}; density of gold \SI{19.3}{g/ml}; atomic weight of gold
197. The volume of a nanostar (assumed spherical) would be
\SI{1E-16}{ml}. This would contain \SI{2E-15}{g} of gold or
\SI{1E-17}{moles}. So \SI{1}{ml} of \SI{0.25}{mM} [Au] should contain
\SI{2.3E10} nanostars.

There would therefore only be enough silver to deposit \emph{just one}
atom on each of \SI{0.0004}{\percent} (about 1 in 260000) of the
nanostars. The rest would have \emph{no} deposited silver. As
mentioned above, such a minimal modification as deposition of a single
silver atom is very unlikely to produce a detectable change of
absorbance of a nanostar; we estimate in the next section the amount of silver deposition necessary to create the spectral changes reported.

Furthermore, at least \SI{99.9996}{\percent} of the nanostars must be
unaltered. They would necessarily have the same spectrum as those in
the zero GOx control. The small admixture of the \SI{0.0004}{\percent}
nanostars each modified by a single silver atom will presumably make
very little difference. Yet hugely different spectra are
reported. Please compare again the black and red spectra, and
consider that the difference is supposed to result from
\SI{0.0004}{\percent} of nanostars having a single silver ion
deposited on them. In reality, a spectrum dominated by the majority
unmodified nanostars and therefore almost identical to the control
spectrum would be expected.

A similar, if slightly less extreme, problem exists for the PSA assays
of Fig.~4, which show very strong signal at \SI{1E-18}{g/ml} PSA and
for which the exact gold concentration is specified (i.e. [Au] =
\SI{0.25}{mM}). If we make the very generous assumption that each PSA
molecule has attached to it 100 GOx molecules, still only about 1 in 4
nanostars will receive a solitary silver ion, with the rest being
unaltered.

\section{Expected blueshift}

I now estimate the amount of silver deposition required to produce the reported spectral shifts of nanostar absorbance.

In general, unadorned gold nanoparticles are associated with a (relatively) red absorbance peak, while those with silver shells display a peak that is closer to the blue. The spectral peaks in Rodriguez-Lorenzo et al.~are rather red-shifted compared to most of the spectra in the literature; presumably because of the relatively large size of the present nanoparticles.

The key observation is that under conditions where silver is supposed to have been deposited on the nanostars, there is no sign of the spectral peak attributable to the unmodified gold nanostars. In particular, the spectrum for \SI{1E-20}{g/ml} GOx of Fig.~2b (red) shows no sign of the peak seen in the control spectra (black and blue). This suggests that the majority of nanostars have been coated with a silver layer sufficient to obscure the gold peak. I'll try to estimate this thickness with reference to work in the literature.

This simple calculation will assume spherical nanoparticles. Conveniently, the densities and atomic weights of silver (\SI{10.3}{g/l} and 108) are such that metallic gold (\SI{19.3}{g/l} and 197) and silver contain very similar numbers of atoms per unit volume.

Kim et al.~\cite{Kim2002} measure spectra before and after silver deposition. They report the spectra of gold-core nanoparticles with silver shells for different mole fractions of the two metals. By a little elementary geometry, we can obtain the thickness $T_\mathrm{Ag}$ of the silver shell from the radius of the gold core ($r_\mathrm{Au}$) and the silver mole fraction ($m_\mathrm{Ag}$):

$$T_\mathrm{Ag} = r_\mathrm{Au}\left(\sqrt[\leftroot{-1}\uproot{2} 3]{(1+m_\mathrm{Ag}(1-m_\mathrm{Ag}))}-1\right).$$

The volume of silver per nanoparticle is

$$V_\mathrm{Ag} = \frac{4}{3}\pi((r_\mathrm{Au} + T_\mathrm{Ag})^3 - r_\mathrm{Au}^3)$$

and if the volume is in cubic metres, the number of silver atoms is then

$$N_\mathrm{Ag} = \frac{\SI{1E7}{}\times 10.3 V_\mathrm{Ag}}{\SI{6E23}{}\times 108}.$$

Fig.~2 of ref \cite{Kim2002} shows the growth of a blueshifted peak that eventually obscures the red peak from the gold core. Two particle sizes of diameters 13\,nm and 25\,nm were tested. With the smaller one, none of the silver mole fractions tested obscured the gold peak in the way seen in Fig.~2b of Rodriguez-Lorenzo et al. Such an effect is, however, observed with the larger particles. The largest silver mole fraction for which the gold peak is still larger than the silver one (and therefore still definitely detectable) is 0.25. This corresponds to an average silver layer thickness of about 1.3\,nm. Even on such small nanoparticles this would imply \SI{1E5} silver atoms per nanoparticle. (Note that the nanostars are larger and have an increased surface area because of their shape, but my aim here is to avoid overestimating the number of silver atoms.)

If there are \SI{2.3E10}{/ml} nanostars (see previous section), that would imply that 1\,ml of solution would require deposition of at least \SI{2.3E15} silver atoms to achieve the observed spectral shift. The discrepancy with the maximum number of \SI{9E4} that could be produced by \SI{1E-20}{g/ml} GOx (calculated above) is a mere factor of \SI{2.5E11}{}. Beside this large number, the various imprecisions in my calculation (size of the nanostars, any specific plasmonic effects associated with the vertices of the nanostars) are probably irrelevant.








\section{Summary}

The premise of inverse sensitivity in Rodriguez-Lorenzo et al.~\cite{Rodriguez-Lorenzo2012},
that a competing reaction can increase the sensitivity of an assay at
the single-molecule limit, seems to be kinetic nonsense. They report
detection of GOx when the reaction volume would only rarely have
contained a single molecule. The detection of small numbers of analyte
molecules does not display the stochastic variability expected. The
detection of tiny numbers of silver atoms is implicitly claimed, but
the assay conditions contain a textbook reaction for producing silver
atoms in large quantities independently of the analyte detection
mechanism. The complete disappearance of the spectral peak of gold
nanostars unmodified by silver atoms is hard to reconcile with the
estimate that only a tiny fraction of stars will receive even a single
silver atom. The apparent discrepancy between the amount of silver
likely to be produced by analyte detection and that estimated to be
required to produce the changes of the absorbance spectrum is a factor
of at least
\SI{2.5E11}{}. The authors should provide a more plausible explanation for their observations.

\section{Acknowledgements}

This analysis is based upon comments I posted on the PubPeer platform as Peer 2:\newline \url{https://pubpeer.com/publications/3E8208F0654769A44C22D4E78DA2B8}.\newline My attention was drawn to the article by the initial comments on the paper by Peer 1.

\newpage \bibliography{molly}

\end{document}